\documentclass[preprint,aps]{revtex4}

\usepackage{graphicx}
\usepackage{dcolumn}

\begin{document}

\preprint{E-mail: lebed@bc.edu}

\title{Magic Angle Effects and AMRO as Dimensional Crossovers}

\author{A.G. Lebed$^{1,2}$, N.N. Bagmet, and M.J. Naughton$^1$}

 \affiliation{$^1$Department of Physics, Boston College, Chestnut Hill, 
MA 02467, USA}

\affiliation{$^2$Landau Institute for Theoretical Physics,
2 Kosygina Street, Moscow, Russia}

\date{March 31, 2004, Submitted to Physical Review Letters}

\begin{abstract}

It is shown that interference effects between velocity 
and density of states, which occur as electrons move 
along open orbits in the extended Brillouin zone, 
result in a change of wave functions dimensionality
at Magic Angle (MA) directions of a 
magnetic field.
In a particular, we demonstrate that these $1D \rightarrow 2D$
dimensional crossovers result in the appearance of sharp 
minima in a resistivity component $\rho_{zz}(H, \alpha)$, 
perpendicular to conducting layers, which explains the main
qualitative features of MA and Angular Magneto-Resistance 
Oscillations (AMRO) phenomena observed in 
low-dimensional conductors (TMTSF)$_2$X, (DMET-TSeF)$_2$X, 
and $\alpha$-(BEDT-TTF)$_2$MHg(SCN)$_4$.

\end{abstract}

\maketitle
  
\pagebreak
  
Low-dimensional organic conductors (TMTSF)$_2$X 
(X = PF$_6$, ClO$_4$, ...), (DMET-TSeF)$_2$X (X = AuCl$_2$, ...),
 and $\alpha$-(BEDT-TTF)$_2$MHg(SCN)$_4$ (M=K, Tl, ...) 
exhibit a number of unconventional angular magnetic oscillations [1-24] 
related to open quasi-one-dimensional (Q1D) sheets of Fermi surface (FS)  
in a metallic phase [1-3],
 \begin{equation}
\epsilon^\pm ({\bf p}) = \pm v_F \ (p_x \mp p_F) - 2 t_b
 \cos(p_y b^*) - 2 t_c \cos(p_z c^*) \  , \ \ \ \ p_F v_F \gg  t_b \gg t_c \ ,
\end{equation}
where $+(-)$ stands for the right (left) sheet of the FS; 
$v_F$ and $p_F$ are the Fermi velocity and Fermi momentum
along conducting ${\bf x}$-axis, respectively; 
$t_b$ and $t_c$ are the overlapping integrals 
between conducting chains; $\hbar \equiv 1$.
Most unconventional angular oscillations in a metallic phase - 
the so-called Danner-Kang-Chaikin oscillations [17], 
the third angular effect [18-20], and the interference commensurate 
(IC) oscillations [20,21] - have been explained in term of Fermi liquid (FL) 
approach to anisotropic Q1D spectrum (1) (see Ref. [17], Ref. [25], 
and Refs. [26,27], correspondingly).

On the other hand, despite the fact that all experimentally observed 
"magic angle" (MA) phenomena [5-16] and AMRO [22-24] are related to 
MA directions [4,28,35] of a magnetic field,
\begin{equation}
\tan \alpha =  (n / m) \ (b^* / c^*)  \ , \ \ \ \ {\bf H} = (0, H \sin \alpha , H \cos \alpha) \ ,
\end{equation}
(where n and m are integers) corresponding to periodic electron 
orbits in $(p_y, p_z)$-plane [4,35], there is no good 
agreement between the numerous theories of MA phenomena [28-39] 
and experiments [5-16] in a metallic phase.
There exist even experimental evidences that, although some MA
effects in a metallic phase [7,16] are of FL origin, the others [3,12-14] 
may significantly break FL picture.
So far, the best qualitative agreements have been achieved between 
the prediction of Ref.[4] and the minima in onset magnetic fields for
field-induced spin-density-wave phases observed at MA directions of 
the field (2) [8,37]. 

The goal of our Letter is to demonstrate that electron wave functions, 
corresponding to open FS in a realistic tight-binding model of Q1D
spectrum with electron hoping only between the neighboring atomic 
sites, 
\begin{equation}
\epsilon^\pm ({\bf p}) = \pm v_x (p_y) \ [p_x \mp p_x (p_y)]  
- 2 t_c \cos(p_z c^*) \ , \ \ \ \ \ p_x (p_y) = p_F + 2 t_b \cos(p_y b^*) / v_F \ ,
\end{equation}
change their dimensionality from $1D$ to $2D$ at MA directions 
of a magnetic field (2) with $m=1$:
\begin{equation}
\tan \alpha = n \  (b^* / c^*)  \ .
\end{equation}
In particular, we show that, in the absence of Landau level quantization 
for open FS (3), the other quantum effects in a magnetic field - Bragg reflections 
result in $1D \rightarrow 2D$ dimensional crossovers at MA
directions of the field (4).

In other words, electron wave functions, which are localized on the 
conducting chains at arbitrary directions of a magnetic field [4,40], 
become $2D$ (i.e., localized on some planes) at the MA directions 
of a magnetic field (4).
As shown below, non-trivial physical origin of these $1D \rightarrow 2D$ 
dimensional crossovers is related to the interference effects between 
velocity component along {\bf z}-axis, $v_z(...)$, perpendicular to conducting  
({\bf x},{\bf y})-planes, and the density of 
states, $v_x (...)$.
These interference effects occur as electrons move along open FS (3) 
in the extended Brillouin zone and are qualitatively different from that
responsible for IC oscillations [27,26].
Using this finding, we demonstrate that it is possible to explain the 
appearance of MA [7,13,15,16] and AMRO [22-24] 
minima in resistivity component 
$\rho_{zz}(H, \alpha)$, perpendicular to conducting planes in 
(TMTSF)$_2$X, (DMET-TSeF)$_2$X, and $\alpha$-(BEDT-TTF)$_2$MHg(SCN)$_4$ 
compounds in the framework 
of FL approach.
We also hope that suggested in this Letter $1D \rightarrow 2D$ 
dimensional crossovers will be the key points in further FL and 
non-Fermi-liquid (n-FL) theories of more 
complex MA phenomena.

At first, let us discuss how $1D \rightarrow 2D$ dimensional crossovers
can lead to the appearance of MA minima in $\rho_{zz}(H, \alpha)$ 
using qualitative arguments.
 For electrons localized on conducting ${\bf x}$-chains [4,40], it is natural to expect
 that conductivity component $\sigma_{zz}(H, \alpha)$ is zero in the absence of 
 impurities  (i.e., at $1/\tau = 0$) and decays as  $1/ \tau^2 \omega^2_c(H) \sim 1/H^2$ 
 at high fields. 
 [Here, $ \omega_c(H)$ is one of the cyclotron frequencies related to electron 
 motion along open FS (3), $\tau$ is an electron relaxation time.]
If, at MA directions of the field (4), electron wave functions become 
delocalized, then $\sigma_{zz}(H, \alpha)$ is expected to have
similarities with conductivity of free electrons at $H=0$. 
Therefore, $\sigma_{zz}(H, \alpha)$ has to saturate at high 
magnetic fields and is expected to be proportional 
to $\tau $.
Below, we demonstrate that this qualitatively different behavior of 
$\sigma_{zz}(H, \alpha)$ at MA directions (4) is indeed responsible for 
the appearance of MA minima 
in $\rho_{zz}(H, \alpha)$.

To develop a quantitative theory, we make use of the Peierls substitution 
method [41] for open electron spectrum [42,4]:
$p_x \rightarrow - i d/dx, \ p_y \rightarrow p_y - (e/c) A_y, \
 p_z \rightarrow p_z - (e/c) A_z $.
 It is convenient to chose vector potential of the magnetic
 field (2) in the form
 ${\bf A} = (0, \ Hx \cos \alpha , \ -Hx \sin \alpha )$, 
 where Hamiltonian (3) in the vicinity of $p_x \simeq p_F$ 
 can be expressed as
\begin{equation}
{\hat \epsilon^{+} ({\bf p})} = v_x \biggl[ p_y b^* - \frac{\omega_b( \alpha ) x}{ v_F}  \biggl] 
\biggl( -i \frac { d }{ d x} - p_x \biggl[ p_y b^* -\frac{ \omega_b( \alpha ) x}{ v_F} \biggl] \biggl) - 2 t_c \cos \biggl[ p_z c^*+ \frac{\omega_c(\alpha) x}{ v_F} \biggl] \ 
\end{equation}
with
\begin{equation}
\omega_b( \alpha ) = e H v_F b^* \cos \alpha / c \ , \ \ \ \ \omega_c( \alpha ) = e H v_F c^* \sin \alpha / c \ 
\end{equation}
being cyclotron frequencies of electron motion along ${\bf y}$-axis and ${\bf z}$-axis,
respectively. 

An important difference between Hamiltonian (5) and the Hamiltonians [27,42,4]
studied so far  is that velocity component along the conducting 
${\bf x}$-chains [i.e., operator of the density of states, ${ \hat v_x(...)}$] 
depends on $p_y$ and $x$.
Although in this case the operators ${\hat v_x(...)}$ and $d/dx$ do not commute, nevertheless one can ignore this fact if the 
quasi-classical parameter 
\begin{equation}
4 t_c / \omega_b( \alpha )  \gg 1 \ .
\end{equation}
It is possible to make sure that, if one represents electron wave functions 
in the form
\begin{equation}
\Psi_{\epsilon} (x, p_y, p_z) = \exp \biggl(
i \int^{x}_{0} p_x \biggl[ p_y b^* - \frac{\omega_b(\alpha) u}{ v_F} \biggl] \ d u \biggl) \
 \psi_{\epsilon}  (x, p_y, p_z) \ ,
\end{equation}
then the solutions of the Schrodinger equation for Hamiltonian (5) can be
written as

\begin{equation}
\psi_{\epsilon} (x, p_y, p_z) = \frac{1}{ \sqrt{v_x[p_y b^* -\frac{\omega_b(\alpha) x}{v_F}]} }  \ \exp \biggl( i \int^{x}_{0} \frac{\epsilon \ du}{v_x [p_y b^* - \frac{\omega_b(\alpha) u}{v_F}]}
\biggl) 
\exp
 \biggl(
 2 i t_c \int^x_0 \frac{ \cos[p_z c^* + \frac{ \omega_c(\alpha) u}{ v_F }]}{v_x[p_y b^* -\frac{\omega_b(\alpha) u}{v_F}]} du \biggl) .
\end{equation}
[In Eq.(9), we normalize wave functions by the standard condition,
 $\int \psi_{\epsilon_1}(x)  \psi_{\epsilon_2}(x) \ dx =  \delta (\epsilon_1 -\epsilon_2)$, 
 and make use of the inequality (7)]. 

Let us demonstrate that suggested in the Letter $1D \rightarrow 2D$ 
dimensional crossovers directly follow from Eq.(9).
It is possible to prove that in the limiting case, where $v_x[...] = v_F = const$,
wave functions (8,9) are always localized on conducting 
${\bf x}$-chains (see Refs. [40,4]).
Below, we show that an account of $p_y$- and $x$-dependences of the
density of states, $v_x [p_y b^* -\omega_b(\alpha) x / v_F]$,
in Eq. (9) lead to de-localization crossovers at MA directions of a
magnetic field (4).
For this purpose, we calculate $z$-dependence of electron wave functions 
at $z=Nc^*$ (where $N$ is an integer) by taking a Fourier transformation 
of the second exponential function in Eq.(9):
\begin{equation}
\Phi (x,p_y,z=Nc^*) = \int_0^{2 \pi} \ \frac{d \ p_z}{2 \pi} \exp(i p_z N c^*) \ 
\exp \biggl(
 2 i t_c \int^x_0 \frac{ \cos[p_z c^* + \frac{ \omega_c(\alpha) u}{ v_F }]}{v_x[p_y b^* -\frac{\omega_b(\alpha) u}{v_F}]} du \biggl) \ .
\end{equation}
After straightforward calculations, $z$-dependence of  electron wave-functions (10) 
can be expressed as
\begin{equation}
\Phi (x,p_y,z=Nc^*) = \exp[-i \phi (x, \alpha, p_y)N] \ 
J_{-N} \biggl[2 t_c \sqrt{I^2_1(x, \alpha,p_y)  + I^2_2(x, \alpha,p_y)} \biggl] \ ,
\end{equation}
where
\begin{eqnarray}
&&I_1 (x = 2 \pi M_0 v_F / \omega_b(\alpha) , \alpha , p_y) =
\sum^{M_0}_{M=0} \int^{2 \pi v_F / \omega_b(\alpha)}_ 0 \ 
\frac{\cos \biggl[ \frac{ \omega_c(\alpha) u}{ v_F} 
+ 2 \pi M \frac{ \omega_c (\alpha)}{ \omega_b(\alpha)} \biggl] 
}{v_x(p_y b^* - \omega_b( \alpha) \ u / v_F)} \ du
\nonumber\\
&&I_2 (x = 2 \pi M_0 v_F / \omega_b(\alpha) , \alpha , p_y) =
\sum^{M_0}_{M=0} \int^{2 \pi v_F / \omega_b(\alpha)}_ 0 \ 
\frac{\sin \biggl[ \frac{ \omega_c(\alpha) u}{ v_F} 
+ 2 \pi M \frac{ \omega_c (\alpha)}{ \omega_b(\alpha)} \biggl] 
}{v_x(p_y b^* - \omega_b( \alpha) \ u / v_F)} \ du \ ,
\end{eqnarray}
with $J_N (...)$ being the Bessel function [43]; $M_0$ 
is an integer.
According to the Bessel functions theory [43], $J_N(Z)$ is an 
oscillatory function of the variable $N$ at $N < |Z|$, whereas
it decays exponentially with $N$ 
at $N > |Z|$. 
Thus, one can conclude that wave functions (10)-(12) are extended along 
$z$-direction if at least one of the functions $I_{i} (...)$ in Eq.(12) 
is not restricted [i.e., if $| I_i(M_0, \alpha ,p_y) | \rightarrow \infty $ 
as $ M_0 \rightarrow \infty $].
In the opposite case, where both functions $I_i(...)$ ($i=1,2$)
are restricted by the conditions $ | I_1 (M_0, \alpha ,p_y) | , 
| I_2 (M_0, \alpha ,p_y) | < I_{max}$, electron wave functions (10)-(12) 
exponentially decay with the variable $z$ at 
$ | z=N c^* | \geq 2 I_{max} $.

Note that functions (12) are written in the form of summations of 
infinite number of electron waves corresponding to electron
quasi-classical motion in different Brillouin zones in the extended
Brillouin zone picture.
Therefore, the physical meaning of summations in Eq.(12) is related
to the interference effects between velocity component along 
${\bf z}$-axis, $v_z = - 2 t_c c^* \sin(p_z c^* + \omega_c ( \alpha) u / v_F)$,
and the density of states, $v_x[p_y b^* - \omega_b ( \alpha) u / v_F]$,
which occur due to Bragg reflections as electron move in a magnetic
field along open orbits.
As it is seen from Eq.(12), angular dependent phase difference between
electron waves, $2 \pi M \omega_c( \alpha) / \omega_b( \alpha)$, is
an integer number of $2 \pi$ only at MA directions (4) of a magnetic field (2),
where $\omega_c( \alpha) = n \ \omega_b( \alpha)$, 
with $n$ being an integer.
Therefore, one comes to the conclusion: at arbitrary direction 
of a magnetic field, the destructive interference effects in Eq.(12) 
result in exponential decay of electron wave functions (10)-(12) 
along ${\bf z}$-axis, whereas, at MA directions, the constructive interference 
effects cause to delocalization of wave functions
along ${\bf z}$-axis.

To calculate conductivity $\sigma_{zz} (H, \alpha)$, let us introduce
the quasi-classical operator of the velocity component $v_z(...)$ 
in a magnetic field [27]:
\begin{equation}
{\hat v_z(p_z,x)} = - v^0_z \sin[p_z c^* + \omega^0_c(\alpha) x / v_F ]  \ , \ \ \ \ v^0_z = 2 t_c c^*.
\end{equation}
Since wave functions (8)-(10) and the velocity operator (13) are known, 
one can calculate $\sigma_{zz} (H, \alpha)$ by means of 
Kubo formalism. As a result, one obtains
\begin{equation}
\sigma_{zz} (H, \alpha) \sim \biggl< 
\frac{1}{v_x(p_y)} \ \int^{0}_{- \infty} \ d (b^* u) \
\frac{\cos[n (\alpha) b^* u]}{\omega_b(p_y +u, \alpha)}
\exp \biggl[ - \int^{0}_{u} \frac{d (b^* u_1)}{\tau \omega_b(p_y +u_1, \alpha)} \biggl] \biggl>_{p_y} \ ,
\end{equation}
where
\begin{equation}
\omega_b (p_y , \alpha) = \omega_b (\alpha) \  [v_x(p_y) / v_F] \ ,
 \ \  \omega_c (p_y, \alpha) = \omega_c (\alpha) \ [v_x(p_y) / v_F] \ ,
 \ \  n ( \alpha) = \omega_c (\alpha) / \omega_b (\alpha) \ ,
\end{equation}
$<...>_{p_y}$ stands for avereging procedure over variable $p_y$.

After straightforward but rather complicated integrations, Eq.(14) can
be rewritten as   
\begin{eqnarray}
\frac{\sigma_{zz} (H, \alpha)}{\sigma_{zz} (0)} =
\frac{1}{1+h^2_c (H)} \ - &&h^2_c(H) \ \int^{0}_{-\infty} \ d u \ \exp(u) \cos[h_c(H) u]
\nonumber\\
&&\times \biggl<  \exp \biggl[ \int^{u}_{0} f[y + u_1 h_b(H)] d u_1 \biggl] - 1 \biggl>_{y}  \ ,
\nonumber\\
f(y) = v_F / v_x(y) -1, \ \ \ &&h_b (H) = \omega_b (\alpha) \tau \ , \ \ \ h_c (H) = \omega_c (\alpha) \tau \ .
\end{eqnarray}
Since in Q1D case $ \rho_{zz}(H, \alpha) \simeq 1 / \sigma_{zz}(H, \alpha)$,
Eq.(16) solves a problem to define $ \rho_{zz}(H, \alpha)$ for electrons with
open spectrum (3) in an inclined magnetic field (2) [44].

To make our results more intuitive, we consider the most important limiting case
- a so-called clean limit, where $\omega_c(\alpha) \ \tau \gg 1$.
In this case, Eq.(16) can be significantly simplified:
\begin{eqnarray}
\frac{\sigma_{zz} (H, \alpha)}{\sigma_{zz} (0)} =&&
\biggl[
\frac{1}{1 + [\omega_c(\alpha) \tau ]^2}  
- \tan^2 \alpha  \  \biggl( \frac{c^*}{2b^*} \biggl)^2  \sum^{\infty}_{n=1} \ \frac{A^2_n}{n^2}
\ \biggl( \frac{2}{1+[\omega_c(\alpha) \tau ]^2}
\nonumber\\
&&-\frac{1}{1+[\omega_c(\alpha) - n \omega_b(\alpha) ]^2 \tau^2}
-\frac{1}{1+[\omega_c(\alpha) + n \omega_b(\alpha)]^2 \tau^2}
\biggl) \biggl]
\end{eqnarray}
where
where $A_n$ are the Fourier coefficients of function 
$f(y) = v_F / v_x(y) -1$:
\begin{equation}
A_N = \frac{1}{ \pi } \int^{+ \pi}_{- \pi} \ f(y) \cos(N y) \ d y \ .
\end{equation}

Eq.(17) directly demonstrates MA maxima in $\sigma_{zz}(H, \alpha)$
[i.e., minima in $\rho_{zz}(H, \alpha)$] related to minima of denominators 
which occur at $ \omega_c(\alpha) = n \omega_b(\alpha)$ 
[i.e., at MA directions of the field (4)].
In Fig.1, we present numerical simulations of Eqs.(17),(18) for three
qualitatively different variants of Q1D spectra (3) corresponding to 
(TMTSF)$_2$PF$_6$, $\alpha$-(BEDT-TTF)$_2$MHg(SCN)$_4$,   
and (TMTSF)$_2$ClO$_4$ 
conductors.
As it is seen, (TMTSF)$_2$PF$_6$ exhibits only one MA minimum, whereas
the last two compounds exhibit several MA minima with large indexes
$n$ in Eq.(4).
We stress that this qualitative feature as well as a doubling of a period of MA 
minima in the case of (TMTSF)$_2$ClO$_4$ are in a good
agreement with the existing experimental data [5,7,13,16,22].

We point out that the existing alternative model to describe
MA and AMRO effects in $\rho_{zz}(H, \alpha)$ - a so-called Osada 
model [30], which is very important from methodological and historical 
points of view, in our opinion, does not have a direct 
physical meaning.
The reason is that the transfer integrals $t_{n,m}$ in Ref.[30] 
are expected to be exponentially small in the framework of 
a realistic tight-binding model [1] of low-dimensional 
electron spectra.
Moreover, as it follows from Eq.(17), a hypothesis [30] that
it is possible to introduce some "effective transfer integrals", $t_{n,m}$, 
in a  linearized electron spectrum (1) and to use such linearized 
spectrum while calculating $\rho_{zz}(H, \alpha)$ 
is incorrect.
Indeed, weighting factors in Eq.(17) depend on magnetic field orientation 
(i.e., on $\tan \alpha$) and, thus, their physical meanings are completely
different from some "effective transfer integrals", $t_{n,m}$, postulated
in Ref.[30].

In conclusion, we hope that $1D \rightarrow 2D$ dimensional crossovers 
suggested in the Letter  will be key points in further theories describing 
more complicated MA phenomena.
In this connection, we point out that there exist three main scenarios 
for  MA phenomena: FL one [4,28-33] based on Gor'kov [42,45] and Chaikin 
[46] approach to Q1D conductors, weak n-FL one [28,35], where electron-electron 
scattering processes depend
on a magnetic field, and n-FL Princeton scenario  [12-14,34,36], where
MA direction of a magnetic field correspond to FL versus 
n-FL crossovers.
Very recently, two novel exotic mechanisms [39,47] have been suggested to
account for MA and AMRO effects.

This work was supported in part by National Science Foundation, grant number
DMR-0076331, the Department of Energy, grant number DE-FG02-02ER63404, and
by the INTAS grants numbers 2001-2212 and 2001-0791.
One of us (AGL) is thankful to N.N. Bagmet and E.V. Brusse for useful discussions.

\pagebreak

\begin{figure}[h]
\includegraphics[width=6.5in,clip]{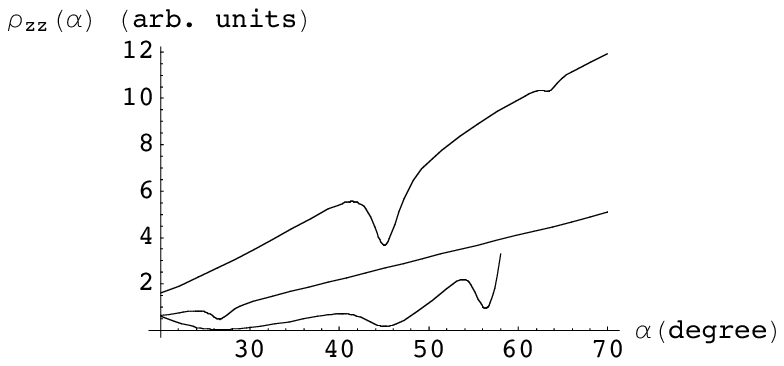}
\caption{ Resistivity component $\rho_{zz}(\alpha) =1/\sigma_{zz}(\alpha)$, 
perpendicular to conducting layers, calculated by means of
Eqs.(17),(18),(5). 
For $\alpha$-(BEDT-TTF)$_2$MHg(SCN)$_4$ (lower curve) and (TMTSF)$_2$PF$_6$
(middle curve) conductors, we use model electron spectrum $\epsilon(p_x,p_y,p_z) = 
2t_a \cos(p_x a/2) + 2t_b \cos(p_y b^*) + 2t_c \cos(p_z c^*)$ [1] with weak, 
$t_a/t_b \simeq 3$, and strong, $t_a/t_b = 8.5$ [27],  Q1D anisotropies, 
respectively.
For (TMTSF)$_2$ClO$_4$ conductor (upper curve), we take into account anion ordering
[1] and, thus, use the spectrum $ \epsilon(p_x,p_y,p_z) = 2t_a \cos(p_x a/2) 
+ \sqrt{[2t_b \cos(p_y b)]^2 + \Delta^2} + 2t_c \cos(p_z c) $ [1]
with $ t_a/t_b = 8.5 $ and $\Delta = 0.2 t_b$. 
In all three cases, we use the value $\omega_b(0) \tau = 15$.}
\label{fig1}
\end{figure}

\pagebreak


\begin{references}
\bibitem{Lebed} As shown in Ref.[35], eigenfunctions of Hamiltonian (1) 
[i.e., Hamiltonian (3),(5) with
 $v_x(...) = v_F = const $] are localized on conducting chains
 at arbitrary inclined magnetic field (2).
\bibitem{Lebed-Lebed} Numerical  simulations of a complex 
three-dimensional integral (16) will be publishes elsewhere.
\end{references}
\end{document}